# RoentMod: A Synthetic Chest X-Ray Modification Model to Identify and Correct Image Interpretation Model Shortcuts

Lauren H. Cooke, Matthias Jung, Jan M. Brendel, Nora M. Kerkovits, Borek Foldyna, Michael T. Lu, and Vineet K. Raghu

## Abstract
Chest radiographs (CXRs) are among the most common tests in medicine. Automated image interpretation may reduce radiologists' workload and expand access to diagnostic expertise. Deep learning multi-task and foundation models have shown strong performance for CXR interpretation but are vulnerable to shortcut learning, where models rely on spurious and off-target correlations rather than clinically relevant features to make decisions. We introduce RoentMod, a counterfactual image editing framework that generates anatomically realistic CXRs with user-specified, synthetic pathology while preserving unrelated anatomical features of the original scan. RoentMod combines an open-source medical image generator (RoentGen) with an image-to-image modification model without requiring retraining. In reader studies with board-certified radiologists and radiology residents, RoentMod-produced images appeared realistic in 93% of cases, correctly incorporated the specified finding in 89-99% of cases, and preserved native anatomy comparable to real follow-up CXRs. Using RoentMod, we demonstrate that state-of-the-art multi-task and foundation models frequently exploit off-target pathology as shortcuts, limiting their specificity. Incorporating RoentMod-generated counterfactual images during training mitigated this vulnerability, improving model discrimination across multiple pathologies by 3-19% AUC in internal validation and by 1-11% for 5 out of 6 tested pathologies in external testing. These findings establish RoentMod as a broadly applicable tool for probing and correcting shortcut learning in medical AI. By enabling controlled counterfactual interventions, RoentMod enhances the robustness and interpretability of CXR interpretation models and provides a generalizable strategy for improving foundation models in medical imaging.

## Main
Deep learning has achieved remarkable success in CXR interpretation, with performance rivaling experienced radiologists[1–4]. Despite this success, clinical adoption of these tools has remained slow because they (i) generalize poorly to out-of-distribution data[5,6] and (ii) arrive at decisions differently than humans and in ways that may be difficult to explain[7,8]. These problems prevent radiologists from effectively collaborating with models[9,10], an essential goal for image interpretation systems[11], and may in fact adversely affect radiologist performance[9].

One potential driver of both issues is shortcut learning, where models use spurious correlations in their training data to arrive at decisions[12]. In medical imaging, there have been several well-established causes of shortcut learning. First, differences in image appearance and outcome prevalence between institutions or care settings within an institution (e.g., ICU vs. outpatients) can lead to models using "markers" (e.g. marker indicating left side of the patient) for these settings as a "shortcut" for the outcome[5,12,13]. Second, models can accurately estimate demographic and body attributes like age, sex, and size from imaging, leading to biased predictions when outcome prevalence is associated with demographics (e.g., cardiomegaly is more common in older adults)[14–16]. Third, the presence of medical devices (e.g., pacemakers, tubes, and lines) can serve as a surrogate for underlying pathology[17], are often easier to identify, and have a greater effect on predictions than the pathology itself[1].

We posit that popular paradigms to train image interpretation models lead to an underappreciated cause of shortcut learning: the use of off-target "tasks" or correlation between outcomes as a surrogate for the outcome itself[18]. Modern models are typically (i) multi-task models trained to predict several derived binary outcomes from text-based reports[19] or (ii) foundation models trained to generate report text itself[20,21]. Often, these models are trained using large observational cohorts with imaging and text reports. This framework is substantially more efficient than training individual models for each outcome[22] and has the added benefit of "feature-sharing"[23] where performance on related tasks improves by simultaneous training. However, multi-task objectives can lead to models encoding image features that are spuriously correlated to multiple outcomes (e.g. supine view indicating sicker patients), as this is often easier than encoding true causal features[18,24].

Current interpretability[25] and shortcut testing tools[14] are useful to identify shortcuts but cannot quantify the impact of shortcuts on model performance or mitigate them directly during training. Counterfactual image editing[26] is a new technique that leverages image generation models to modify an existing image via a text-based prompt. In this way, a pathological finding can be "added" or "removed" from an image (see Figure 1) while keeping other aspects of the image fixed[27,28]. Current counterfactual CXR editing tools either require expensive hardware[28,29], lack radiologist expert review and validation[27–29], or depend on components that are not publicly available[28].

Here, we introduce RoentMod, a free, open-source, counterfactual image editing tool to add a pathological finding to an existing CXR image. Through experiments on large publicly available datasets and multitask diagnostic models, we make the following contributions:

- We present an accurate counterfactual medical image editing tool (RoentMod) developed by combining a medical image generation model (RoentGen)[30] with a publicly available image modification model built for natural images (Image-to-Image)[31] with no additional training of either model.
- We conduct an evaluation study with radiologists, showing that RoentMod accurately modifies real chest radiographs to produce counterfactual radiographs that a) appear realistic, b) accurately add a user-specified finding, and c) largely maintain the other anatomy of the original radiograph.
- We show that using RoentMod to add pathology to a CXR results in increased probability predictions for all tested pathologies (including ones that were not added to the CXR), indicating that state of the art multitask diagnostic models may use shortcut learning.
- We show that a novel training paradigm that combines real scans with their RoentMod-modified counterfactual CXRs in a mini-batch reduces the use of these shortcuts and improves discrimination performance across all pathologies in internal testing on the NIH-CXR 14 and MIMIC-CXR Datasets and external testing on PadChest and CheXpert.

## Results

### Approach Overview

We aimed to develop a generative model to modify CXRs based on a user-specified text prompt. Our model, RoentMod, combines the CXR knowledge encoded in the weights of the RoentGen[32] CXR generation model with the Stable Diffusion Image-to-Image architecture[31] for image modification. RoentGen[32] is the Stable Diffusion text-to-image diffusion model [31] finetuned on MIMIC-CXR to produce synthetic CXR images that adhere to a user-specified text prompt.  We first assessed the quality of RoentMod synthetic scans (Figure 2) across three dimensions: 1) realism and adherence to the text prompt (e.g., adding a right upper lobe lung mass when prompting "right upper lobe mass"), 2) limited addition of unrelated conditions (e.g., adding hernia when prompting "right upper lobe mass"), 3) subject-identity preservation (maintaining the individual's anatomy outside of the prompted pathology). Second, we used RoentMod to evaluate how existing CXR interpretation models' predictions change with the addition of off-target pathology (e.g. adding cardiomegaly when predicting pneumonia) by comparing predictions across pairs of real unmodified scans with "normal" chest x-rays (i.e. no finding noted in the radiology report) and their RoentMod synthetic counterparts after adding pathology. We interrogated four pretrained models in this work: two models from the torchxrayvision[33] library (Torch X-ray trained on all cohorts named txrv-all, Torch X-ray trained on the NIH CXR-14 cohort only named txrv-nih), the ElixrB CXR foundation model[34], and the Ark+ CXR foundation model[20] (Supplemental Table 3). We performed this analysis on sets of seven scans per patient: one original unmodified scan and six RoentMod-modified synthetic versions, where each modified scan contained a preselected pathology. This allowed us to assess how the addition of one pathology affects the model's predictions for all other pathologies. Finally, we trained our own diagnostic models on synthetic and real scans to attempt to remove reliance on shortcuts.

### RoentMod Evaluation

**RoentMod accurately adds pathology from text prompts while maintaining realism.** One radiology resident (4 years of experience) and one board-certified radiologist (8 years of experience) individually assessed a total of 800 (400 each) RoentMod-modified scans blinded to input prompts. Across all tested conditions, RoentMod maintained realistic subject appearance in 745 out of 800 (~93%) synthetic scans. Radiologists observed notable pathology outside of the eight tested in this study in only 20 out of 800 (~4%) scans (Supplemental Figure 2). Of our eight tested conditions, RoentMod consistently introduced six pathologies when prompted: 92% of cases for cardiomegaly, 97% for edema, 93% for pleural effusion, 89% for pneumonia, 99% for hernia, and 99% for lung mass. When prompted to generate emphysema or lung nodules, RoentMod only added these conditions in 76% and 70% of cases, respectively; however, these results were inconsistent across readers (Supplemental Figure 3).

**RoentMod modifies images to add prompted pathology without introducing medically unrelated off-target conditions.** RoentMod mostly generated images without including medically unrelated unprompted conditions (Supplemental Figure 2). Cardiomegaly frequently appeared when RoentMod was prompted to add edema (92% of cases), which is reasonable pathophysiology as pulmonary edema is usually related to poor cardiac function and often accompanied by cardiomegaly. Cardiomegaly also appeared frequently in RoentMod images of all other tested conditions (35-50% of cases per prompted pathology). RoentMod also tended to add edema when

prompted to add cardiomegaly (46%), emphysema (33%), and lung nodules (50%). Based on these findings and the relatively lower prevalence of emphysema and nodules when RoentMod was prompted to include them (76 and 70%), we elected to remove nodules and emphysema from the rest of the study.

**RoentMod preserves the subject's other anatomy when adding a pathology.** To determine whether RoentMod maintains subject identity when adding pathologies, we measured pairwise Fréchet inception distance[35] (pFID) across the InceptionV3[36], XResNet[37], and CLIP[38] embeddings between baseline real scans with RoentMod synthetic scans from a different person (control score), baseline real scans with RoentMod synthetic scans for the same person (model score), and baseline real scans with real follow-up scans with the added pathology having naturally occurred within two years of the corresponding baseline scan (real score) (Supplemental Table 2). We found that synthetic images were more similar to their baseline real scan than two real CXRs from different individuals. The choice of embedding determined whether synthetic CXRs or real follow up CXRs were more similar to baseline CXRs from the same person. Using CLIP embeddings, we found that synthetic CXRs were more similar than follow-up CXRs to real images. However, we observed the opposite trend for InceptionV3 embeddings (>10 points per condition) and XresNet embeddings (within 5 points for all conditions except 31 points for hernia and 10 for lung masses.

## Using RoentMod to add pathology to CXRs results in higher predicted probability of other pathologies

We then used RoentMod to "stress-test" existing multitask and foundation x-ray interpretation models (Supplemental Table 3). To do this, we assessed the change in predicted probability between pairs of original images from the MIMIC-CXR cohort without pathological findings and their RoentMod-modified counterparts after adding a pathology (Figure 3 Panel A). We then identified potential shortcuts by assessing the change in predicted probability percentile for all tested pathologies when adding a single pathology using RoentMod.

We found that all tested models likely exhibit shortcut learning, as adding a single off-target pathology changed the diagnostic model's prediction probabilities for all other tested pathologies (e.g. adding edema changes the prediction of hernia) (Figure 3 Panel A; off-diagonal entries). Txrv-all exhibited strong changes in predicted probability for all findings when RoentMod added a lung mass, pleural effusion, cardiomegaly, or pulmonary edema. The multi-task model trained only on the NIH cohort had less evidence of shortcut learning based on change in predicted probability when adding an off target finding. The ElixrB model demonstrated strong shortcut learning for all pathologies tested, especially when predicting pneumonia or pulmonary edema (percentile change from 58 to 75%). The Ark+ foundation model had the smallest use of shortcut learning but still used presence of a lung mass as a shortcut for all findings (percentile increase from 20 to 58%). Both foundation models exhibited a strong tendency to identify pulmonary edema on counterfactual scans regardless of the added condition. We found similar results when using baseline scans from NIH CXR-14 (Supplemental Figure 4).

## Using RoentMod scans during training reduces reliance on shortcuts and improves generalization

We then trained a multi-task model using both real scans from NIH CXR-14 and RoentMod-generated counterfactual scans to mitigate shortcut learning. We first compared discrimination (area under the ROC curve, or AUC) on held-out NIH CXR-14 (in-distribution), MIMIC-CXR (RoentGen-training cohort), CheXpert (out of distribution), and PadChest (out of distribution) testing sets. We found that incorporating RoentMod-modified scans generally improves diagnostic accuracy beyond a multi-task model trained only on real NIH CXRs (Torch X-ray NIH). We found a particularly large AUC improvement (3-19%) in NIH CXR-14 and MIMIC-CXR (Table 1). Notably, we found that our RoentMod trained multitask diagnostic model's in-distribution AUC surpassed zero-shot predictions from foundation models trained on much larger cohorts. We further note that all tested models besides ElixrB were trained on NIH CXR-14 with different train-test splits, so reported AUCs are likely overestimates. In contrast, in out of distribution data (PadChest and MIMIC-CXR), we find that models trained on much larger cohorts outperform our RoentMod-trained multitask model.

We then tested our RoentMod-trained multitask model using the same counterfactual stress test approach (Figure 3 Panel B). We found that our model largely corrects shortcut learning on counterfactual scans for all tested conditions, except the model still overrepresents presence of pneumonia on scans with added lung masses and cardiomegaly on scans with added pulmonary edema when compared to the radiologist-read co-occurrence rates of these pathologies as ground truth. These two co-occurrences do however make clinical sense, as pneumonia is common in lung cancer[39] and the typical presentation of edema is cardiogenic. Of note, due to the counterfactual scans only adding one finding at a time, the model trained using these scans tends to exhibit negative shortcut learning, underpredicting findings when another is present. We found similar results when using NIH CXR-14 baseline scans.

## Discussion

Chest radiographs (CXRs) are among the most common tests in medicine[40] and accurate, automated interpretation is poised to greatly reduce radiologists' burden [11,20]. Since many pathologic findings may appear on CXR, multi-task and foundation models are more efficient and can be more accurate than training specialized models for each task. However, they are also more vulnerable to shortcut learning, where models rely on spurious correlations to make decisions. Here, we introduced RoentMod to identify and reduce shortcut learning. RoentMod generates anatomically realistic counterfactual CXRs with synthetic pathology while preserving the original features on the image. Using RoentMod, we show that multi-task interpretation models and foundation models use off-target pathology as shortcuts and that incorporating RoentMod-generated images during training removes this shortcut and can improve interpretation models.

Our major findings are:
- An accurate counterfactual medical image editing tool can be developed by combining a medical image generation tool (RoentGen[32]) with the publicly available image modification tool (image-to-image[31]) with no additional training of either model.

- In user studies with board-certified radiologists, RoentMod accurately modifies real chest radiographs to produce synthetic radiographs that a) appear realistic 93% of the time, b) accurately add a user-specified finding 89-99% of the time, and c) maintain the anatomy of the original radiograph akin to real follow-up scans.
- Many publicly available multi-task and foundation models use off-target pathologic findings as shortcuts, suggesting that these models lack specificity to individual pathology.
- A training paradigm that incorporates RoentMod counterfactual images together with real images corrects this shortcut learning and improves discrimination compared to a model trained on the same real dataset in internal (3-19% AUC improvement) and external testing (improvement on 5/6 pathologies).

Several studies have reported shortcut learning by image interpretation models including CXR models[12–14]. These have largely focused on protected demographic attributes like age, sex, and racial identity or image acquisition/technical parameters[5]. Here, we use RoentMod's counterfactual image generation to show that both conventional multi-task learning and foundation model training lead to models that use off-target pathologic findings as "shortcuts." (i.e., generally sicker individuals have higher predictions for all pathologic findings). We show that vetting the counterfactual image generation tool through radiologist evaluation and then incorporating counterfactual images during training can mitigate shortcut learning and lead to more generalizable models with minimal training data.

Counterfactual medical image generation is growing in popularity with several tools recently made available[27–29]. RoentMod distinguishes itself through its ease of use and computational efficiency. RoentMod is a combination of publicly available image modification and CXR generation models and required no additional training to generate anatomically realistic counterfactual CXRs. These CXRs are generated in <2 seconds/image on consumer GPUs (NVIDIA RTX 4090) and <30 seconds/image on a standard CPU. Additionally, we conducted a thorough evaluation of RoentMod using both radiologist and in silico studies and found that RoentMod generates anatomically realistic radiographs that adhere to a user-specified text prompt and preserve the subject's anatomy.

While our study focused on mitigating shortcuts based on off-target pathology in CXR image interpretation models, RoentMod and other counterfactual medical image generation tools could be used in numerous other medical imaging applications. Examples include fairness evaluations to ensure that demographic changes do not result in substantially different interpretation performance and stress-testing and fine-tuning segmentation models to ensure that anatomic segmentation is unbiased based on the presence/absence of pathology. Currently, a major focus of AI research is on foundation models and automated report generation from medical images[20,22,34]. Here, we show that zero-shot classification performance of a foundation model suffers from the same shortcuts as multi-task interpretation models, suggesting that counterfactual images may improve report generation tools as well. This is especially important in medical imaging, where achieving the massive scale of other foundation models (e.g., large language models) is impractical due to data regulation, patient privacy, and limited ability to acquire additional data.

Limitations of our study should be considered. In this work, we focused on synthetically modifying CXRs to add eight common pathologies; however, a major benefit of foundation models is that they can simultaneously identify many pathologies. As with RoentGen[32], RoentMod requires clearly defined pathologies (e.g., "multiple pulmonary nodules" vs. "nodule") with short context in user text prompts. Though we observed robust performance in diverse external validation on the tested pathologies, fine-tuning RoentMod on more diverse datasets beyond the CXRs from a single institutional database in Boston (MIMIC-CXR) may improve the diversity of synthetically generated images and improve upon existing realism errors as given in Supplemental Figure 2. Even though radiologists reported high realism scores for RoentMod CXRs, we did not include real images in the reader studies as a negative control, potentially leading to optimistic results. Here, we focused on CXR, but we expect that interpretation models for other imaging modalities will likely suffer from similar shortcuts; this will be tested in future work. We also observed that co-occurrence rates between some pathologies were higher than expected in synthetic scans (e.g., cardiomegaly often appeared regardless of the user prompt). Additional techniques like structural causal models or region-specific editing could be used to improve RoentMod and further mitigate these correlations in future work. Our model trained using RoentMod counterfactual images exhibited a slight "reverse" shortcut learning since we generated images with only one pathology. Generating multiple pathologies on a single image or recalibration during training may be necessary to prevent this and may result in improved generalization performance. We also note that our RoentMod-trained diagnostic model only had exposure to real NIH-CXR 14 scans. Although performant on this limited dataset in comparison to its multitask diagnostic model counterparts, we hypothesize that a multitask diagnostic model trained on more examples across more cohorts with this RoentMod-supplemented training paradigm could potentially perform at or better than these larger models. Lastly, subtle changes may be present in counterfactually generated CXRs that are not visible to human readers, affecting our ability to accurately perform shortcut testing.

In summary, we introduced RoentMod, a counterfactual CXR image generation tool to identify and mitigate shortcut learning in CXR interpretation models. RoentMod generates realistic counterfactual images that adhere to a text prompt without changing unrelated anatomy. Using RoentMod, we find that multi-task and foundation models use the presence of any pathology as a shortcut to interpret CXRs. Finally, using real and synthetic CXRs during training removes this shortcut and improves generalization performance of multi-task interpretation models.

## Methods

This work relies solely on publicly available datasets and models. Readers who have completed the necessary access requirements can use all datasets in the study. All resources produced in this study, including the RoentMod synthetic scan modification tool, a sample dataset of synthetic scans, all statistical analysis code, and radiologist reads of synthetic scans are publicly available to use for further research.

### Datasets

We collected and used three de-identified chest x-ray (CXR) datasets with associated pathology labels. All pathology co-occurrence rates and pathology frequency by cohort are given in Supplemental Figure 1 and Supplemental Table 1.

**National Institutes of Health (NIH) Chest X-ray 14**[41]**.** This cohort consists of 112,120 inpatient frontal radiographs from 30,805 patients collected between 1992 and 2015 at the NIH Clinical Center in Bethesda, MD[42]. We only used the posterior-anterior (PA) scans taken on patients aged 18 or older, for a final cohort of 64,628 scans over 27,713 patients (mean age 47.8 ± 15.0 years, 47% female). For each image fourteen text-mined disease labels were extracted from the corresponding scan's radiology report into a database: atelectasis, consolidation, infiltration, pneumothorax, edema, emphysema, fibrosis, pleural effusion, pneumonia, pleural thickening, cardiomegaly, nodule, mass, and hernia. Entries without any findings from this list were labeled no finding.

**MIMIC-CXR**[43]**.** This cohort consists of 227,835 Chest X-Ray imaging studies (scan and radiology report) from 65,379 patients collected between 2011 and 2016 at the Beth Israel Deaconess Medical Center in Boston, MA. We limit our work to frontal scans and patients aged 18 or older, for a final cohort of 94,067 scans over 44,642 patients (mean age 56.1 ± 19.0 years, 53% female). We then used the labels that MIMIC-CXR produced on their dataset with the CheXpert labeler[44], where each image contains fourteen text-mined pathology presence labels from the corresponding scan's radiology report: no finding, atelectasis, consolidation, pneumothorax, edema, pleural effusion, pneumonia, pleural other, cardiomegaly, lung lesion, lung opacity, enlarged cardio mediastinum, fracture, and support devices. Any values labeled as uncertain were treated as negative labels for presence of that condition.

**CheXpert**[44]**.** This cohort consists of 224,316 inpatient and outpatient radiographs from 65,240 patients collected between 2002 and 2017 at Stanford Hospital in Palo Alto, CA. We only used PA radiographs for patients 18 or older, for a final cohort of 29,453 scans over 20,574 patients (mean age 57.1 ± 17.7 years, 38% female). Each image contains fourteen text-mined pathology presence labels from the corresponding scan's radiology report: no finding, atelectasis, consolidation, pneumothorax, edema, pleural effusion, pneumonia, pleural other, cardiomegaly, lung lesion, lung opacity, enlarged cardio mediastinum, fracture, and support devices. Any values labeled as uncertain were treated as negative labels for presence of that condition.

**PadChest**[45]**.** This cohort consists of 168,861 Chest X-Ray imaging studies (scan and radiology report) from 67,625 patients collected between 2009 and 2017 at the *Hospital Universitario de San Juan* in Alicante, Spain. We limit our work to frontal scans and patients aged 18 or older, for a final cohort of 88,109 scans over 59,085 patients (mean age 58.6 ± 17.4 years, 52% female). We used labels extracted from the radiology reports as described by the PadChest curators. [45]

**Related Work**

Shortcut learning in medical image interpretation is the tendency for deep learning models to use spurious correlations in training data instead of the actual presentation of the pathology to make decisions. Most techniques to identify shortcut learning have focused on model interpretability more broadly. These fall into three main categories: heatmap-based techniques, shortcut testing, and counterfactual image editing.

Heatmap-based approaches (e.g., saliency maps[46], Grad-CAM[47], and its variants)[48,49] aim to find areas in the image that most strongly affect a model's output. These techniques assign an importance value to each pixel or region of an image to produce a heat map of influence for a model's prediction[50]. These importance values typically correspond to the gradient of the prediction with respect to the pixel value (i.e., how much the output would change if the pixel value changed). This approach has been used to successfully identify visually apparent shortcuts in multitask medical imaging diagnostic models[5]; however, saliency maps require experts to manually review large sets of maps to identify shortcuts[13]. More concerningly, recent studies have found saliency maps can remain nearly identical after randomizing model weights, suggesting that saliency maps capture dataset-specific patterns (e.g., edges in an image) rather than a model's particular decision-making process[25,51,52].

Shortcut testing techniques aim to directly test whether trained models rely on user-specified shortcuts by investigating models' latent representations. Recent work has shown that models tend to have reduced prediction depth (layer at which a prediction is finalized) when using an obvious shortcut[53]. Others have relied on the assumption that spatial features are less likely to be shortcuts than pixel intensity-based representations[54] to identify shortcuts and improve fairness[12,55]. Another technique, ShorT[14], assumes that shortcuts will reduce fairness metrics across a hypothesized shortcut attribute and tests whether model-based encoding of the attribute reduces fairness for the real target. For both heatmap-based and direct shortcut testing, it is unclear how to mitigate shortcuts once identified. Image augmentation techniques appear to improve fairness[55,56] but likely do not prevent all shortcuts.

Counterfactual generation techniques create new input images that remove or alter image features to directly investigate shortcuts. The first set of these techniques (e.g. Gifsplanation, Generative Visual Rationales)[57,58] use Generative Adversarial Networks (GANs) and Variational Autoencoders (VAEs) to produce counterfactual images most similar to the original image but with a different model prediction. While GAN-based methods have successfully found realistic features in models trained to predict heart failure in cardiac MR[59] and CXR[60], they also produce lower resolution generations and suffer from unstable training regimes[61].

Modern approaches use Stable Diffusion's open-source image generation models[31] and fine-tuned medical image versions (e.g., RoentGen[32]) to find shortcuts using text-guided, high-quality medical imaging counterfactuals. BioMedJourneys[29] fine-tuned Stable Diffusion with GPT-4 based instruction fine-tuning to generate image counterfactuals from a baseline image and associated text describing the changes. This approach accurately produces counterfactual images but suffers from hallucinations and a lack of radiologist-based synthetic image evaluation. RadEdit[28] introduced a diffusion-based generative model to edit chest radiographs using human-drawn image masks to restrict editing to relevant areas while minimizing unrelated artifacts (e.g., introduction of support devices). RadEdit can produce counterfactual images through image editing; however, it is not known whether RadEdit can identify shortcuts or mitigate identified model biases. Further, it is difficult to apply RadEdit at scale because it requires input bounding boxes for each image. PRISM[27] reconsiders this image editing approach by removing the need for image masks by fine-tuning the Stable Diffusion Text-to-Image[31] model's denoising U-Net directly to edit CXRs and permit higher-resolution image generation (512x512 pixels). Using PRISM-generated images during training improves classifier performance for four pathologies. However, PRISM can only edit findings documented in CheXpert[44], and the effectiveness of PRISM's generated images was evaluated using classifier prediction changes, which may not reflect radiologist interpretation. A set of novel approaches aim to use structural causal models to ensure that counterfactual images avoid spurious correlations during generation[62,63]. While promising, these approaches require accurate description of the data generating process including all confounders, and more work is required to generate high-resolution medical images efficiently.

Building on previous image editing work, RoentMod leverages existing models instead of retraining from scratch. We create a diffusion-based image editor by combining RoentGen's pretrained U-Net and text encoder weights with pretrained weights from Stable Diffusion's Image-to-Image model[31], requiring no additional training. Crucially, we collaborate with board-certified radiologists to validate RoentMod, rather than relying on model-based validation. We can then use generated images not only to reveal shortcut learning in interpretation models but also incorporate synthetic images during training to mitigating these shortcuts. This work aims to make AI-based medical imaging tools more trustworthy in correcting these shortcuts, taking an important step toward bridging the gap between research and clinical use for these tools.

### RoentMod development and design

**RoentMod Design: No Training Required.** RoentMod allows a user to input an existing CXR and a text prompt to get a new scan of that individual that reflects changes requested by the prompt (Figure 1). To build RoentMod, we applied the pretrained weights from RoentGen[32] to the Stable Diffusion Image-to-Image architecture[31]. We used the finetuned weights for the text encoder and Denoising U-net from RoentGen and the default weights from the Stable Diffusion Image-to-Image architecture's variational autoencoder. This approach required no additional training or fine-tuning beyond these existing models.

**Parameter and Prompt Selection.** The Stable Diffusion Image-to-Image architecture uses two main hyperparameters to control the output image: guidance scale and strength. Higher guidance controls better adherence to the text prompt often at the expense of lower image quality. Strength controls the extent to which the model can alter the reference image, with 0 meaning no change and 1 meaning maximal change (through additional added noise prior to denoising). We optimized guidance scale and strength across tested conditions using five randomly selected images with no pathologic finding. We generated all tested conditions for all combinations of ten values for guidance scale evenly spaced between 1.5 and 10, and ten values for strength evenly spaced between .2 and 1 (Supplemental Figure 5). We had a board-certified radiologist review and notate which parameter combinations produced the most realistic images that contained the requested prompt. From this process, we selected RoentMod's guidance scale and strength parameters to be 0.4 and 4 respectively for all experiments in this work.

### RoentMod Evaluation

**Synthetic Evaluation Cohort.** To evaluate RoentMod's performance, we first generated a set of synthetic scans by modifying 100 randomly selected scans with no documented pathologic finding from MIMIC-CXR (see Datasets). Per real CXR based on our prompt selection work from before, we had RoentMod produce a synthetic CXR for each of the eight pathology-generating prompts (Supplemental Table 4; 800 synthetic CXRs in total).

Using this synthetic dataset, we tested whether RoentMod can correctly modify existing CXRs with realistic new pathology and preserve unrelated anatomical features of the input individual. We measured RoentMod's performance along three main metrics: 1) realism and adherence to the text prompt, 2) limited addition of unrelated conditions, and 3) subject-identity preservation.

**Evaluating RoentMod: Realism and Adherence to the text prompt.** We first assessed RoentMod's ability to realistically produce the prompted pathology when modifying scans. Two board-certified radiologists read all synthetic scans for 100 randomly selected MIMIC-CXR patients in the synthetic cohort for a total of 800 scans read (400 randomly assigned scans per person). We asked each radiologist (Supplemental Table 4) to label presence/absence/unsure of each of the eight findings investigated in the study for each image. We also allowed radiologists to write open-ended notes in reference to image realism and adding extra anomalies. We then manually converted these notes into binary scores for artificialness (1 if artificial), and extra anomaly addition (1 if extra additions present).

We then assessed the co-occurrence between the radiologist-read labels and the labels according to the specified text prompt to RoentMod (Supplemental Figure 2). In this matrix, each cell represents the fraction of scans that have the prompted pathology specified in the rows and the radiologist labelled pathology specified in the columns. Thus, the diagonal of the co-occurrence matrix represents adherence to the text prompt.

**Evaluating RoentMod: limiting unrelated conditions.** We then measured how often RoentMod produces unprompted pathology. Using the same co-occurrence matrix (see Methods, Adherence to the text prompt), we assessed the fraction of scans where we prompt RoentMod to produce one condition, but radiologists found another condition on the synthetic scan (i.e., off-diagonal elements of the matrix). We indexed the co-occurrence of RoentMod generated scans with that of the full real NIH CXR-14 cohort as a surrogate for true co-occurrence.

**Evaluating RoentMod: subject-identity preservation.** Finally, we evaluated whether RoentMod-modified scans are anatomically similar to the real input scans using pairwise Fréchet Inception Distance (FID)[35] across the InceptionV3[36], XResNet[37], and CLIP[38] image embeddings. FID traditionally measures the digitally embedded similarity between two sets of images in the InceptionV3[36] image embedding space, however here we only include one image in each set to find the similarity directly between pairs of images and permit evaluation using different embeddings (we call pFID). This work uses pFID to evaluate image similarity between real scans with no pathologic finding paired with a RoentMod synthetic scan from a different person, real scans with no pathologic finding paired with a RoentMod synthetic scan from the same person, and real scans with no pathologic finding paired with a real follow-up scan for the same person when that individual had naturally developed the pathology under investigation within 2 years of the first scan. To generate paired baseline and follow-up similarity scores, we identified pairs across the entire NIH dataset where these follow up events occurred, giving us 3261 evaluated pairs across all categories over 2986 total individuals. For patients who had multiple eligible pairs for a given condition, we randomly selected one pair to represent that individual. We list the number of evaluated pairs per condition along with FID results in the appendix (Supplemental Table 2). We then assessed similarity between real and synthetic scans and similarity between two real scans from different individuals using all 2986 of these individuals. We generated synthetic scans of each of our six evaluated pathologies per person, for a total of 2986 pairs of scans per pathology.

## Using RoentMod to evaluate existing image interpretation models

Existing multitask CXR diagnostic models aim to read an input CXR and output one probability score per pathology reflecting the model's belief that the CXR contains the pathology. We used RoentMod to "stress-test" these diagnostic models. For each real CXR with no pathologic finding, we generated a counterfactual CXR with one added pathology. This allows us to directly measure whether models' predictions are impacted by off-target pathology. Based on our evaluation of RoentMod, we determined that synthetic scans were sufficient quality to evaluate the impact of six pathologies: cardiomegaly, edema, pleural effusion, pneumonia, hernia, and pulmonary mass.

**Pretrained models.** We interrogated four multitask pretrained CXR diagnostic models where model weights were publicly available in this work: two densenet121[64] architecture models from the torchxrayvision[33] library, each trained on different cohorts (Torch X-ray trained on all cohorts called txrv-all, Torch X-ray trained on the NIH CXR-14 cohort only called txrv-nih), the ElixrB CXR foundation model[34], and the Ark+ CXR foundation model[20]. Supplemental Table 3 describes each model's training datasets, architecture, and diseases that the model can predict.

**Measuring change in predicted probability.** We measured the change in predicted probability for each condition of interest by comparing outputs on pairs of (i) real CXR with no pathologic finding from MIMIC-CXR or NIH CXR-14 with (ii) synthetic CXR with synthetically added pathology. To control for variable calibration of the tested models, we then converted each output probability to a percentile score (between 0-100%) reflecting how the probability compares to the predictions of the model on the entire cohort. Finally, we calculated the median change in percentile score between the baseline scan and after a pathology was counterfactually added to the scan (Supplemental Figure 4). We separately report the median predicted probability on the baseline and modified scans compared to the true pathology co-occurrence from the radiologist reads (see Methods Evaluating RoentMod adherence to the text prompt) as a reference standard (Supplemental Figure 6).

## Using RoentMod as data augmentation to improve interpretation models

Lastly, we investigated if we could use RoentMod scans as data augmentation to reduce reliance on shortcuts and improve overall performance of multi-task CXR interpretation models.

**Synthetic Training Cohort.** To train our own classifier and avoid data leakage in the process, we generated 10,000 images with no pathologic finding using RoentGen's[30] CXR generator ("no acute cardiopulmonary process"). We then had RoentMod produce two synthetic CXRs for each of the original eight pathology-generating prompts (Supplemental Table 4). Overall, RoentMod produced a total of 160,000 synthetic scans for this cohort based on the 10,000 RoentGen baselines for a total of 170,000 synthetic scans.

**Model Training.** We first experimented with single-task or multi-task, the ratio of real to synthetic scans during training, and how off-target pathology should be labeled on counterfactual CXRs. We selected our baseline parameters through a series of small-scale training experiments on 2000 and then 5000 synthetic patients. We trained Densenet121 models to predict each pathology in single- and multi-task setups. Since it is not clear whether RoentMod synthetic scans contain off-target pathology (e.g., prompted cardiomegaly often contains edema), we experimented with different labelling schemes for pathology not mentioned in the text prompt including (i) assuming off-target pathologies were not present, (ii) ignoring off-target pathologies, or (iii) using target values equal to the radiologist-read co-occurrence values (Supplemental Figure 2). We generated baseline patients with no pathologic finding using RoentGen's[30] CXR generator ("no acute cardiopulmonary process") and then used RoentMod to generate counterfactual scans using the same process as generating the synthetic cohort (see Synthetic Training Cohort) but for the 6 final pathologies, creating a total of seven scans per patient (one no finding, six with findings – one per pathology). After training, we then evaluated each of these small models across the full NIH CXR-14 cohort to produce AUC scores per model per pathology (Supplemental Figure 7).

From this work we chose to train a multi-task model on a randomly selected set of 20,000 patients from the NIH CXR-14 cohort (65315 scans) combined with the training portion of the synthetic training cohort of 136,000 scans. All remaining scans from NIH CXR-14 were reserved for testing. We used co-occurrence values from the RoentMod radiologist reads (Supplemental Figure 2) as the ground truth for off-target pathology for RoentMod synthetic scans (e.g., to label whether pleural effusion is present on a scan that we prompted to add cardiomegaly). We trained the model with a learning rate of 0.0001 for 100 epochs, with a batch size of 32 and early stopping enabled after 50 epochs without improvement.

**Evaluation of models trained with RoentMod data augmentation.** We then assessed whether models trained with RoentMod augmented data had (i) reduced sensitivity to off-target pathology and (ii) had improved in-distribution and out-of-distribution AUC. We evaluated whether the trained model with RoentMod augmented data was sensitive to addition of off-target pathology on the full NIH CXR-14 and MIMIC-CXR cohorts for their available labels using the same process as for the pretrained models (Supplemental Table 3). Lastly, we calculated the AUC of our model and all tested pretrained models across held-out NIH CXR-14 participants, the full MIMIC-CXR dataset, and the full PadChest dataset to assess whether data augmentation with RoentMod improves performance in in-distribution, in the generator's training distribution, and in out-of-distribution datasets, respectively.

## Data Availability
Pre-existing cohorts (MIMIC-CXR, NIH-CXR 14, and PadChest) are publicly available after completing each cohort's access requirements. Our synthetic cohort will be made available for research by request upon publication.

## Code Availability
All models and code will be available upon publication.

## Acknowledgements
We gratefully acknowledge the contributions of the teams behind several open-access datasets and tools that made this work possible. We want to thank the NIH Clinical center for their work on Chest-Xray 14, the MIT Lab for Computational Physiology for MIMIC-CXR, The University of Alicante Machine Learning and Medicine Lab for PadChest, and the Stanford Machine Learning Group for their work on the CheXpert. These annotated CXR datasets with standardized labels have been instrumental in advancing automated radiograph interpretation. We also want to thank the developers of the TorchXRayVision library for their set of open source pretrained models and methods that made our experimental process standardized and easier to reproduce. We also would like to thank Hugging Face and Stability AI for their model hosting and open access image-to-image diffusion model architecture respectively. Finally, we extend our appreciation to the creators of RoentGen for their foundational work in generative medical imaging, which served as a conceptual and practical reference for our methods.

## Competing Interests
**Matthias Jung** is funded by the Deutsche Forschungsgemeinschaft (DFG, German Research Foundation) #518480401
**Jan M Brendel** is funded by the Deutsche Forschungsgemeinschaft (DFG, German Research Foundation) #540505270
**Michael T Lu** reports funding to his institution from the American Heart Association, Amgen, AstraZeneca, Ionis, Johnson & Johnson Innovation, Kowa Pharmaceuticals America, MedImmune, National Academy of Medicine, National Heart, Lung, and Blood Institute, and Risk Management Foundation of the Harvard Medical Institutions outside the submitted work
**Borek Foldyna** reports institutional research support from NIH/NHLBI, AstraZeneca, MedImmune, Cleerly, and MedTrace, all outside the submitted work.
**Vineet K Raghu** is funded by AHA Career Development Award 935176 and NHLBI K01HL168231.

## Main Figures and Tables

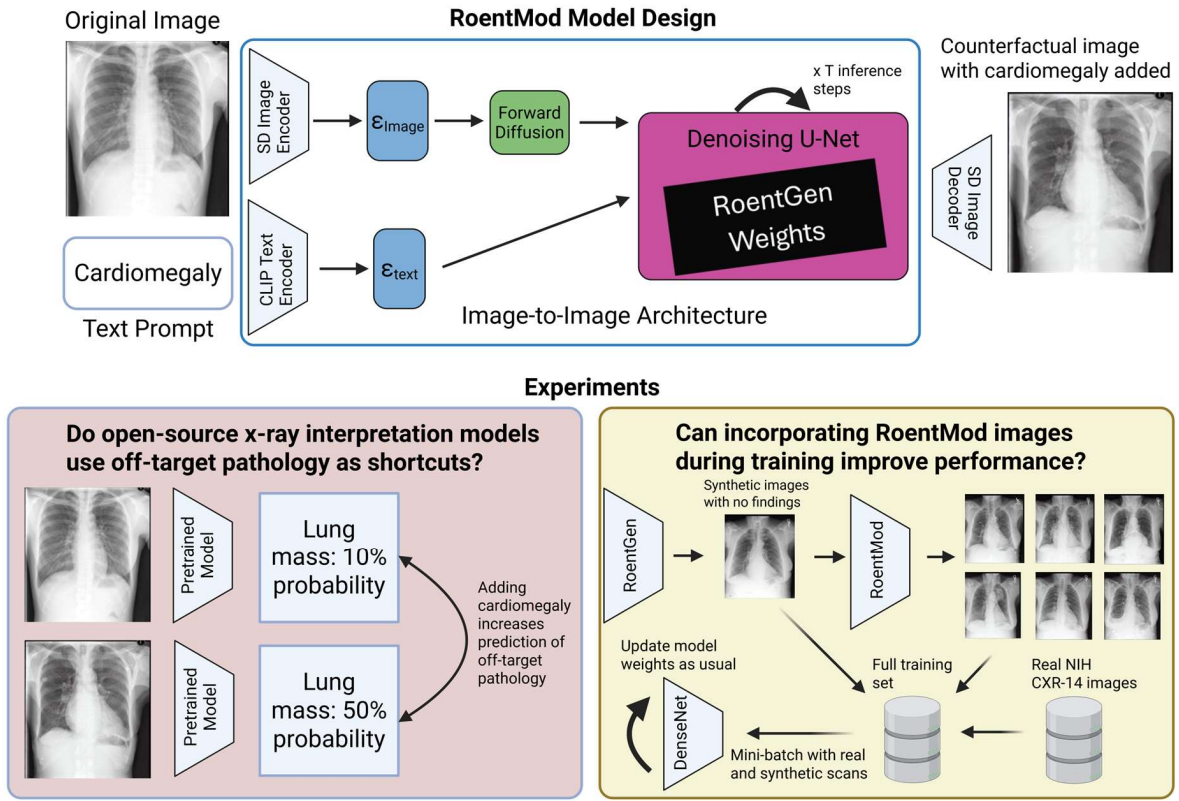

Figure 1: Graphical Abstract

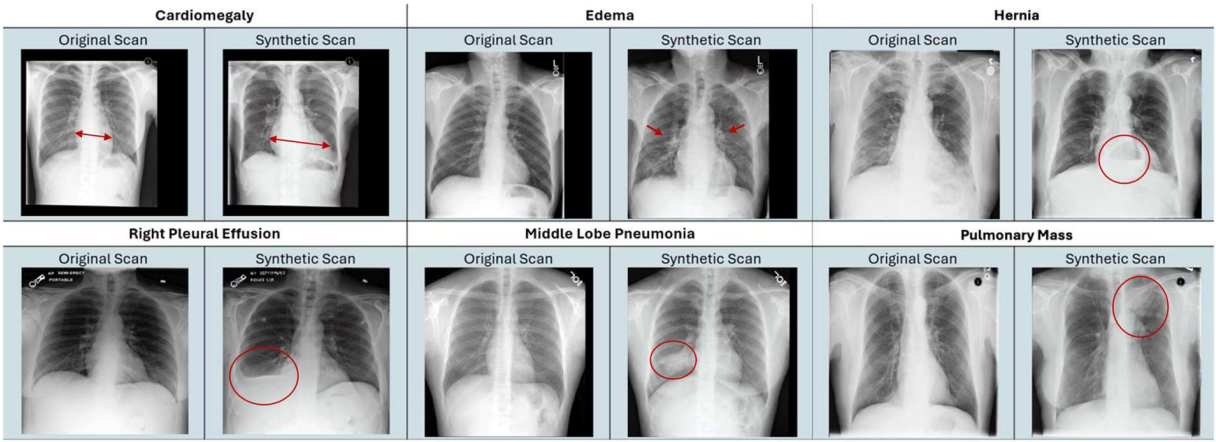

Figure 2: Representative real and counterfactual annotated chest radiographs produced by RoentMod

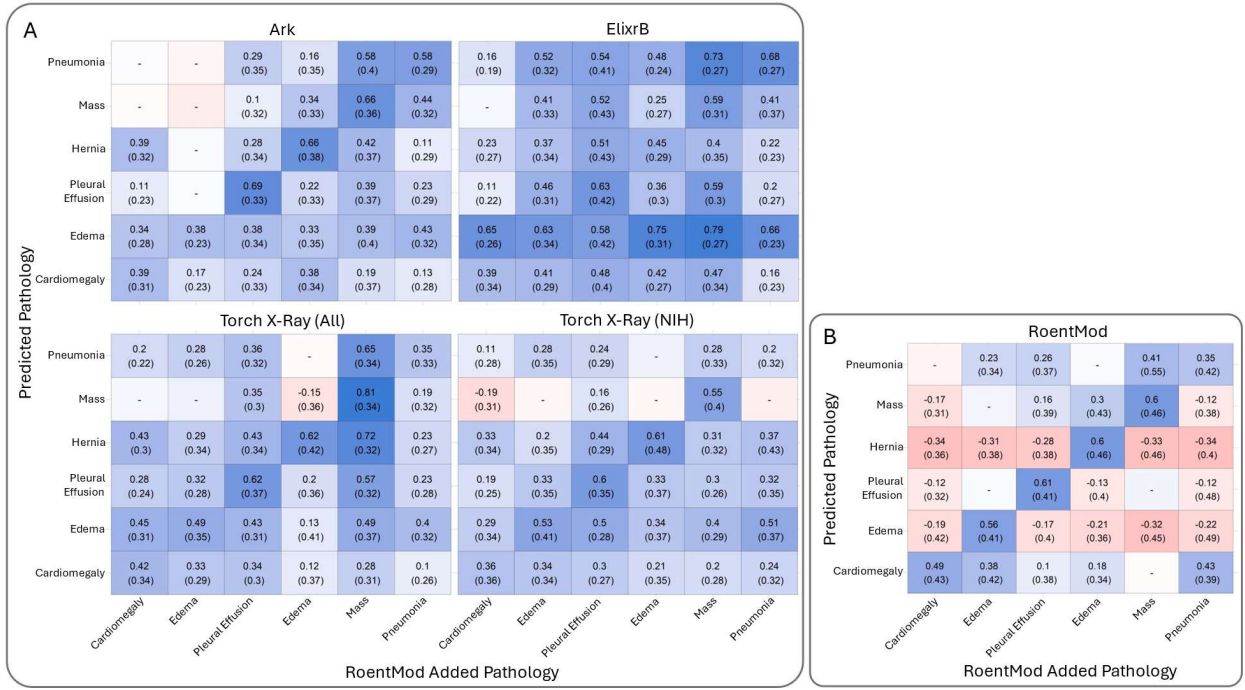

Figure 3: Effect of adding pathologies on predicted probabilities from existing multitask CXR interpretation models (Panel A) and our RoentMod-trained multitask CXR interpretation model (Panel B). Blue boxes indicate greater sensitivity to adding pathology. Blank boxes indicate no change between baseline and RoentMod-generated counterfactuals. Values reflect the change in predicted probability percentile after pathology is added to scans with no finding in MIMIC-CXR

| NIH AUC | Cardiomegaly | Edema | Pleural Effusion | Pneumonia | Hernia | Mass |
|---|---|---|---|---|---|---|
| **RoentMod (NIH)** | **0.96** | **0.96** | **0.94** | **0.80** | **0.99** | 0.89 |
| Torch-Xray (NIH) | 0.93 | 0.77 | 0.89 | 0.74 | 0.88 | 0.83 |
| Torch-Xray (All) | 0.91 | 0.82 | 0.90 | 0.73 | 0.97 | 0.85 |
| Ark | 0.94 | 0.90 | 0.90 | 0.79 | 0.93 | **0.91** |
| ElixrB | 0.82 | 0.88 | 0.89 | 0.73 | 0.78 | 0.82 |
| **MIMIC AUC** | Cardiomegaly | Edema | Pleural Effusion | Pneumonia | Hernia | Mass |
| **RoentMod (NIH)** | 0.81 | 0.77 | 0.88 | 0.55 | 0.89 | 0.61 |
| Torch-Xray (NIH) | 0.76 | 0.69 | 0.82 | 0.53 | 0.68 | 0.63 |
| Torch-Xray (All) | 0.84 | 0.79 | 0.87 | 0.56 | **0.91** | 0.65 |
| Ark | **0.86** | **0.83** | **0.93** | **0.61** | 0.86 | **0.75** |
| ElixrB | 0.81 | 0.80 | 0.91 | 0.56 | 0.71 | 0.63 |
| **Padchest AUC** | Cardiomegaly | Edema | Pleural Effusion | Pneumonia | Hernia | Mass |
| **RoentMod (NIH)** | 0.91 | 0.81 | 0.85 | 0.69 | 0.91 | 0.85 |
| Torch-Xray (NIH) | 0.89 | 0.78 | 0.84 | 0.67 | 0.80 | 0.86 |
| Torch-Xray (All) | **0.93** | 0.81 | 0.89 | 0.75 | **0.96** | 0.91 |
| Ark | **0.93** | **0.88** | **0.94** | **0.88** | 0.87 | **0.94** |
| ElixrB | 0.89 | 0.85 | 0.89 | 0.79 | 0.73 | 0.87 |
| **CheXpert AUC** | Cardiomegaly | Edema | Pleural Effusion | Pneumonia | Hernia | Mass |
| **RoentMod (NIH)** | 0.81 | 0.77 | 0.86 | 0.57 | - | 0.59 |
| Torch-Xray (NIH) | 0.74 | 0.72 | 0.83 | 0.54 | - | 0.58 |
| Torch-Xray (All) | 0.81 | 0.84 | 0.89 | 0.67 | - | 0.65 |
| Ark | **0.88** | **0.91** | **0.93** | **0.76** | - | **0.73** |
| ElixrB | 0.79 | 0.80 | 0.89 | 0.67 | - | 0.61 |

Table 1: Discrimination (AUC) of publicly available and RoentMod-trained multi-task models across diverse cohorts

# Supplemental Figures and Tables

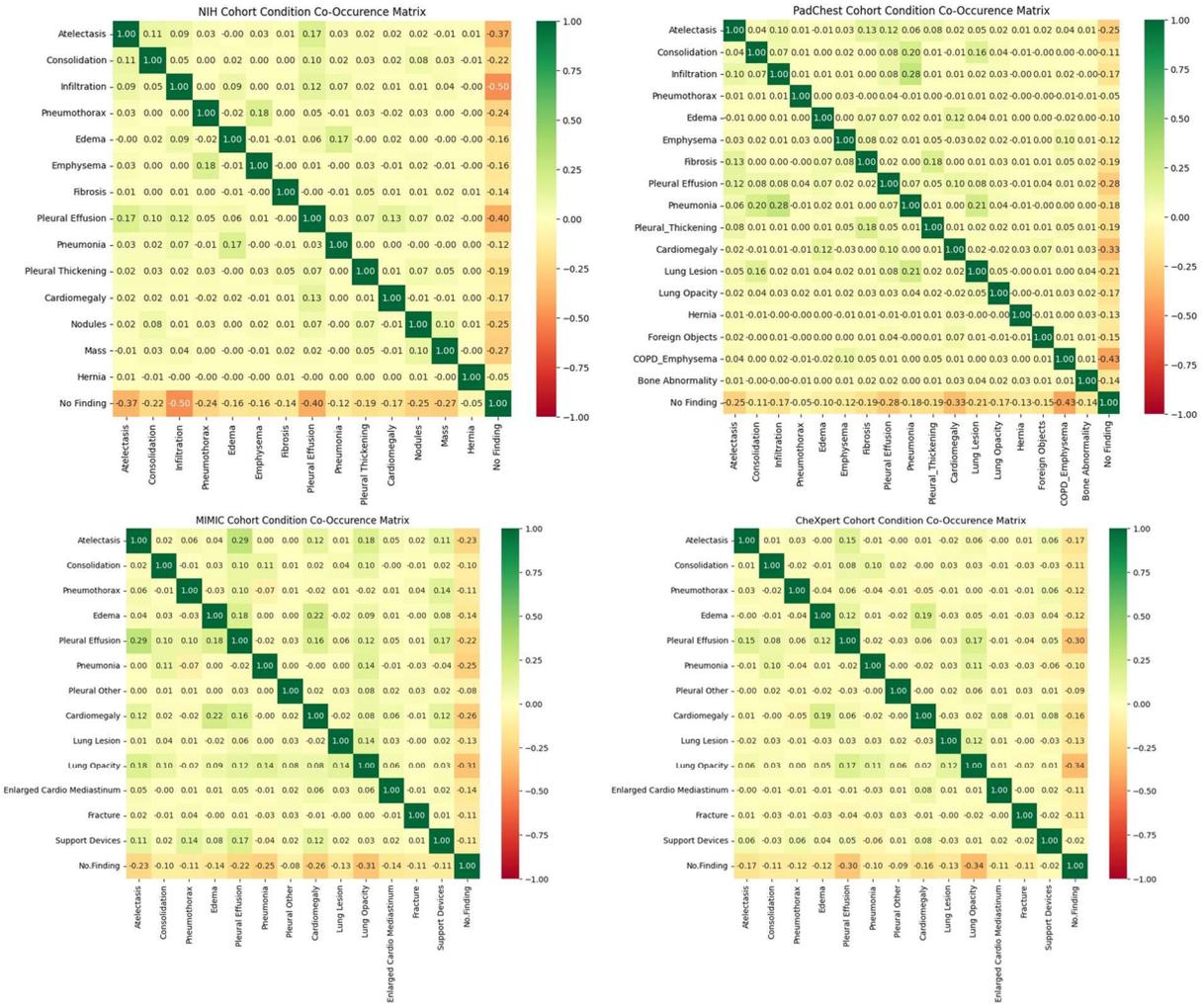

Supplemental Figure 1: Co-occurrence of findings in NIH, MIMIC, CheXpert, and PadChest cohorts

| Counts | NIH CXR-14 (64628 scans) N = 27713 | MIMIC-CXR (94067 scans) N = 44642 | CheXpert (29453 scans) N = 20574 | PadChest (88109 scans) N = 59085 |
|---|---|---|---|---|
| Mean Age (stdev) | 47.8 (15.0) | 56.1 (19.0) | 57.1 (17.7) | 58.6 (17.4) |
| % Female (N female) | 46% (12759) | 53.0% (23679) | 38% (7805) | 52.4% (30982) |
| Atelectasis % (# scans) | 8.7% (5614) | 16.9% (15912) | 10.9% (3199) | 5.9% (5199) |
| Consolidation % (# scans) | 2.3% (1463) | 4.0% (3745) | 5.1% (1498) | 1.2% (1072) |
| Infiltration % (# scans) | 13.9% (8976) | -- | -- | 2.9% (2563) |
| Pneumothorax % (# scans) | 5.1% (3268) | 4.5% (4220) | 6.1% (1804) | 0.2% (207) |
| **Edema % (# scans)** | 0.4% (268) | 7.1% (6671) | 5.8% (1710) | 0.9% (791) |
| **Emphysema % (# scans)** | 2.3% (1473) | -- | -- | 1.3% (1187) |
| Fibrosis % (# scans) | 2.2% (1394) | -- | -- | 3.3% (2935) |
| **Pleural Effusion % (# scans)** | 10.0% (6450) | 16.1% (15152) | 27.4% (8081) | 7.1% (6240) |
| **Pneumonia % (# scans)** | 0.9% (586) | 19.8% (18589) | 4.1% (1198) | 3.2% (2789) |
| Pleural Thickening % (# scans) | 3.7% (2366) | -- | -- | 3.6% (3213) |
| Pleural Other % (# scans) | -- | 2.5% (2390) | 3.4% (1012) | 0.5% (433) |
| **Cardiomegaly % (# scans)** | 2.4% (1520) | 21.1% (19844) | 9.9% (2910) | 9.9% (8680) |
| **Lung Lesion % (# scans)** | 5.4% (3466) | 6.0% (5670) | 7.2% (2122) | 4.2% (3705) |
| Lung Opacity % (# scans) | 6.4% (4106) | 26.9% (25300) | 33.1% (9744) | 2.9% (2538) |
| Enlarged Cardio mediastinum % (# scans) | -- | 7.2% (6758) | 4.9% (1438) | 0.7% (631) |
| Fracture % (# scans) | -- | 4.5% (4262) | 4.8% (1411) | 4.3% (3762) |
| **Hernia % (# scans)** | 0.3% (191) | -- | -- | 1.7% (1481) |
| Support Devices % (# scans) | -- | 16.7% (15738) | 27.7% (8146) | 4.5% (3944) |
| **No Finding % (# scans)** | 58.0% (37452) | 20.8% (19561) | 18.7% (5513) | 50.2% (44259) |

Supplemental Table 1: Characteristics of NIH CXR 14, MIMIC-CXR, CheXpert, and PadChest cohorts

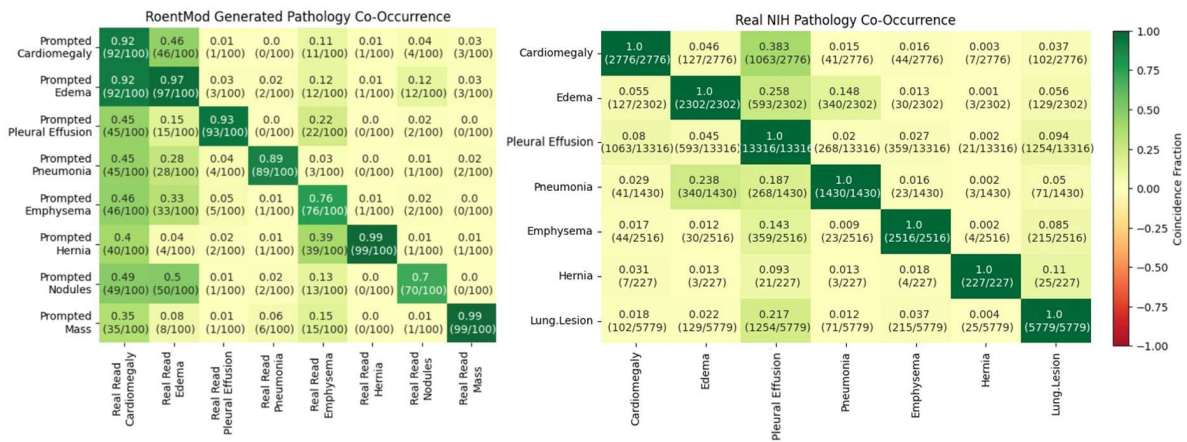

Supplemental Figure 2: Co-occurrence between prompted pathology and radiologist read pathology in RoentMod generated counterfactual CXRs (right) vs. observed co-occurrence between pathologies in NIH CXR-14 (left)

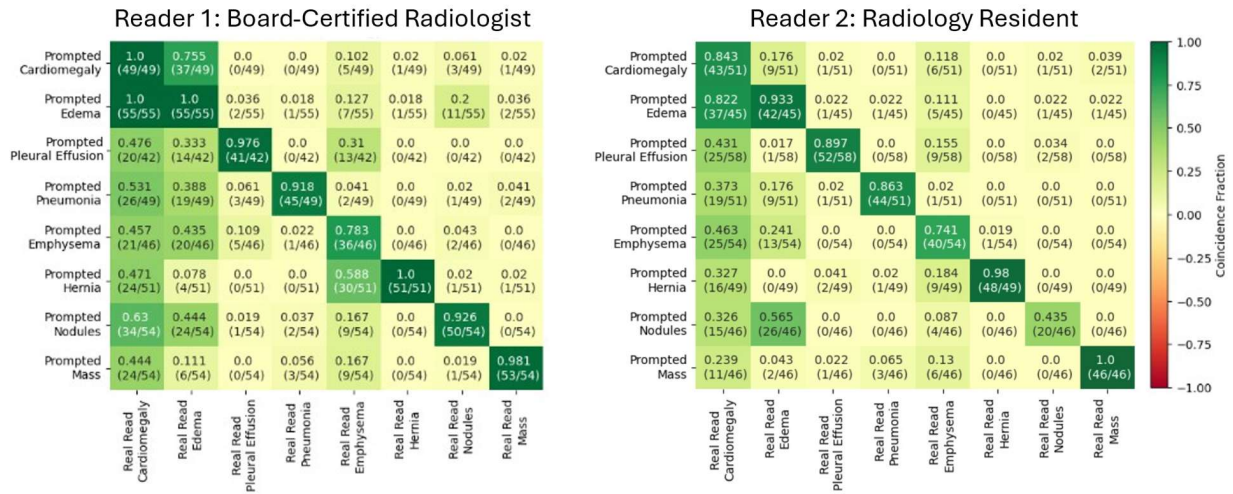

Supplemental Figure 3: Comparison between radiologist-evaluated reads (columns) vs. RoentMod-prompted pathology (rows) stratified by reader

| FID Median (IQR) (N=2986) | | Cardiomegaly N=395 | Edema N=167 | Pneumonia N=302 | Pleural Effusion N=1764 | Hernia N=38 | Pulmonary Mass N=897 |
|---|---|---|---|---|---|---|---|
| InceptionV3 Embeddings | Control Score | 211.05 (73.8) | 222.99 (75.9) | 205.12 (73.0) | 202.96 (71.2) | 223.97 (74.3) | 202.79 (72.2) |
| | Real Score (≤ 2 yrs follow up) | **138.09 (66.78)** | **143.85 (65.9)** | **138.81 (60.1)** | **141.55 (71.9)** | **125.44 (51.9)** | **133.30 (72.0)** |
| | Model Score | 156.53 (60.48) | 184.68 (69.2) | 147.70 (61.5) | 156.73 (63.5) | 182.52 (62.3) | 159.68 (64.0) |
| XResNet Embeddings | Control Score | 196.05 (80.66) | 195.42 (78.3) | 190.30 (81.9) | 182.67 (74.3) | 201.12 (76.5) | 190.21 (74.3) |
| | Real Score (≤ 2 yrs follow up) | 119.07 (63.68) | **129.10 (69.5)** | **114.31 (59.8)** | 121.24 (63.2) | **116.90 (53.9)** | **113.21 (66.4)** |
| | Model Score | **118.99 (52.93)** | 134.35 (57.6) | 114.80 (61.5) | **121.01 (51.9)** | 147.04 (57.1) | 123.31 (52.7) |
| CLIP Embeddings | Control Score | 0.1138 (0.184) | 0.1094 (0.175) | 0.1161 (0.196) | 0.1168 (0.196) | 0.1101 (0.177) | 0.1207 (0.200) |
| | Real Score (≤ 2 yrs follow up) | 0.0825 (0.169) | 0.0608 (0.108) | 0.0797 (0.214) | 0.0726 (0.170) | 0.0848 (0.201) | 0.0747 (0.175) |
| | Model Score | **0.0091 (0.158)** | **0.0149 (0.022)** | **0.0077 (0.014)** | **0.0094 (0.019)** | **0.0150 (0.025)** | **0.0108 (0.196)** |

Supplemental Table 2: Fréchet Inception distance between unpaired synthetic-real scans (control score), paired synthetic-real scans (model score), and real scans with real follow-up scans (real score) across embeddings (rows) and conditions (columns)

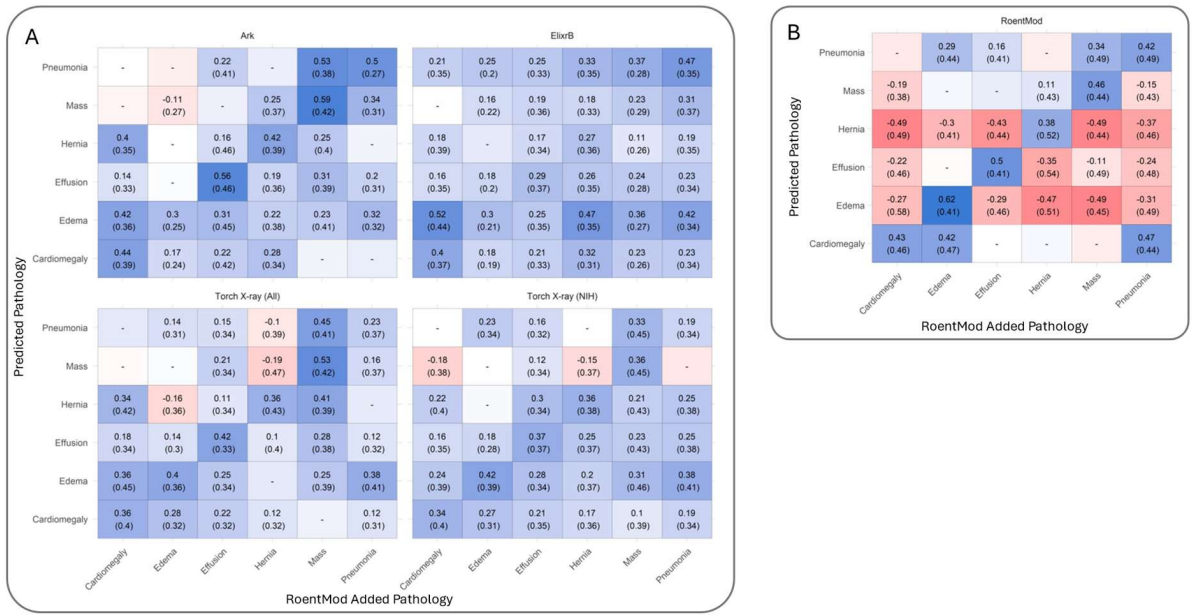

Supplemental Figure 4: Effect of adding pathologies on predicted probabilities from existing multitask CXR interpretation models (Panel A) and our RoentMod-trained multitask CXR interpretation model (Panel B). Blue boxes indicate greater sensitivity to adding pathology. Blank boxes indicate no change between baseline and RoentMod-generated counterfactuals. Values reflect the change in predicted probability percentile after pathology is added to scans with no finding in NIH-CXR 14.

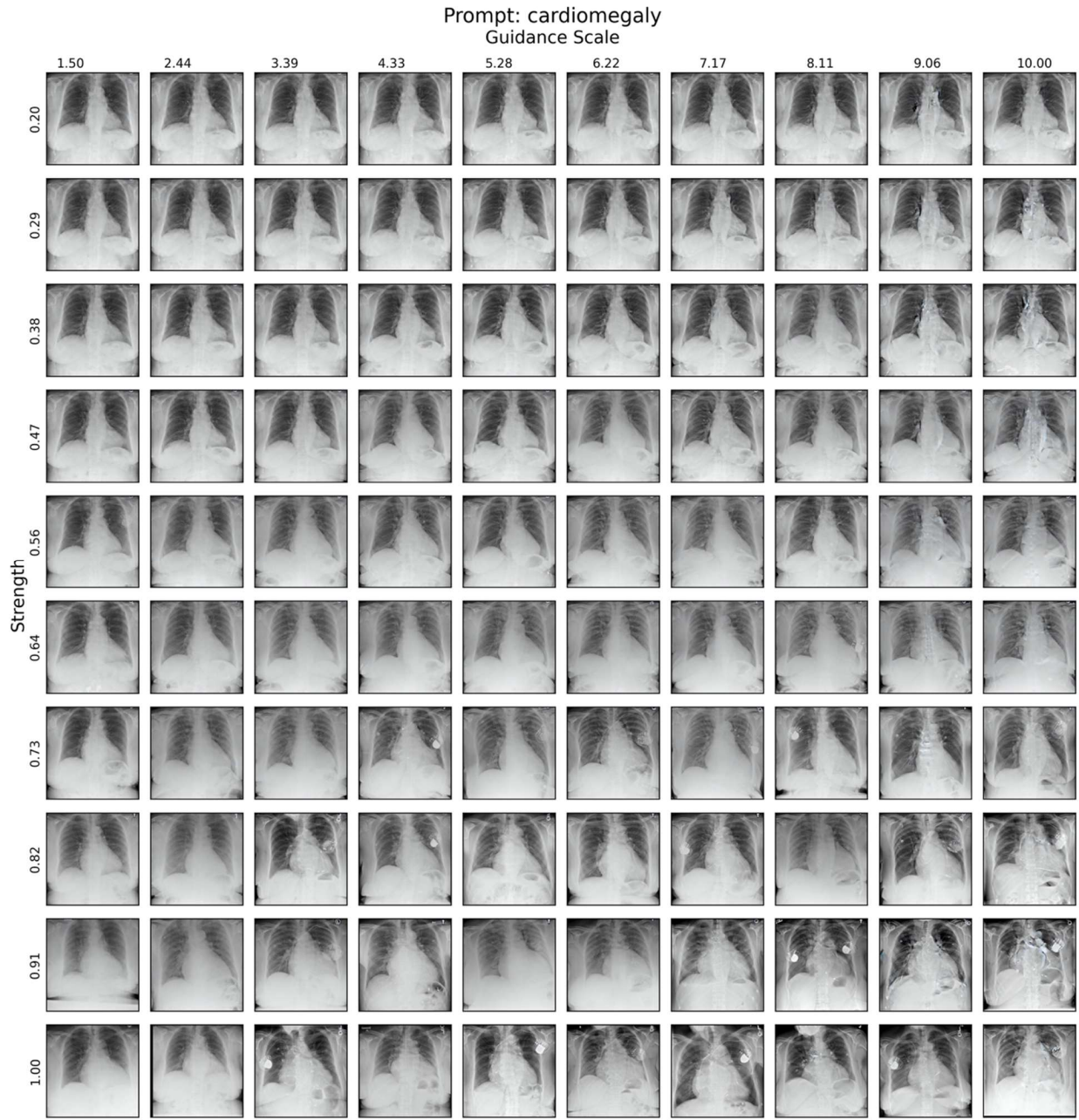

Supplemental Figure 5: Effect of strength and guidance on RoentMod-generated chest radiographs

| Model Name | Model Type | Training Datasets | Model Architecture | Tested Model Predictions | Code Reference |
|---|---|---|---|---|---|
| txrv-all | Multitask classification | NIH, MIMIC-CXR, CheXpert, PadChest | Densenet121<br>224px input image resolution | Cardiomegaly, Edema, Pleural Effusion, Mass, Pneumonia, Hernia | https://github.com/mlmed/torchxrayvision?tab=readme-ov-file |
| txrv-nih | Multitask classification | NIH | Densenet121<br>224px input image resolution | Cardiomegaly, Edema, Pleural Effusion, Mass, Pneumonia, Hernia | https://github.com/mlmed/torchxrayvision?tab=readme-ov-file |
| ElixrB (Google) | Foundation | MIMIC-CXR, 485,082 proprietary CXRs from India, 165,182 proprietary CXRs from a hospital in Illinois, USA | EfficientNet-L2 CNN image encoder and a BERT-based text encoder, trained jointly via supervised contrastive, CLIP, and BLIP-2 losses<br>1024px input image resolution | Cardiomegaly, Edema, Pleural Effusion, Mass, Pneumonia, Hernia | https://huggingface.co/google/cxr-foundation |
| Ark+ | Foundation | MIMIC-CXR, NIH, CheXpert, RSNA Pneumonia, VinDr-CXR, Shenzhen | Swin-Transformer-based image encoder with a momentum teacher-student architecture and projection head<br>1024px input image resolution | Cardiomegaly, Edema, Pleural Effusion, Mass, Pneumonia, Hernia | https://github.com/jianglab/Ark |

Supplemental Table 3: Publicly available chest radiograph interpretation models evaluated in this study

| Target Pathology | Tested Prompts | Final Prompts |
|---|---|---|
| No finding | "no acute cardiopulmonary process" | "no acute cardiopulmonary process" |
| Cardiomegaly | "cardiomegaly" | "cardiomegaly" |
| Edema | "edema", "butterfly edema" | "edema" |
| Pneumonia | "pneumonia", "right upper lobe pneumonia", "left upper lobe pneumonia", "middle lobe pneumonia", "right lower lobe pneumonia", "left lower lobe pneumonia" | "middle lobe pneumonia" |
| Pleural Effusion | "right pleural effusion", "left pleural effusion", "right pleural effusion" | "right pleural effusion" |
| Emphysema | "emphysema", "severe emphysema", "panlobular emphysema" | -- |
| Hernia | "hernia" | "hernia" |
| Pulmonary Nodules | "solitary lung nodule", "multiple pulmonary nodules" | -- |
| Pulmonary Mass | "right upper lobe mass", "left upper lobe mass", "middle lobe mass", "right lower lobe mass", "left lower lobe mass" | "left upper lobe mass" |
| **Reader Instructions** | | |
| For each of the following anonymized scans in your assigned set, please locate the scan number corresponding to the scan ID column (ie. row with scan ID 9 for scan 9.jpg). For each row, fill out each column to indicate if the specified condition is present (1), not present (0 or blank), or unsure (2). (ie. if scan x has Cardiomegaly and nothing else, put a 1 in the cardiomegaly column for that row and nothing everywhere else) If anything about the particular image is odd or you need to note something beyond these labels, please type that information into the corresponding row's Notes column. | | |

Supplemental Table 4: RoentMod Prompts (original and pruned) and reader instructions

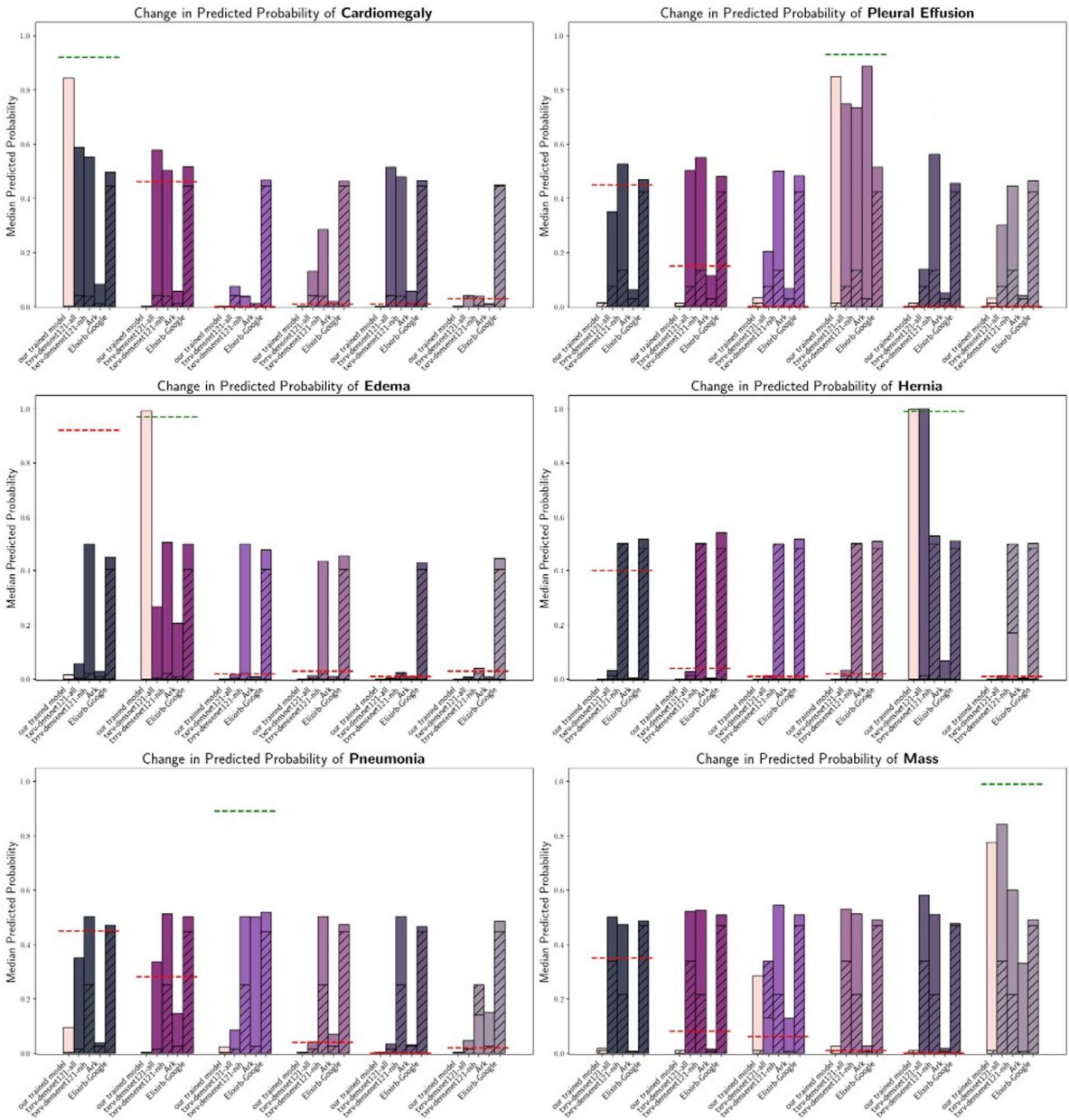
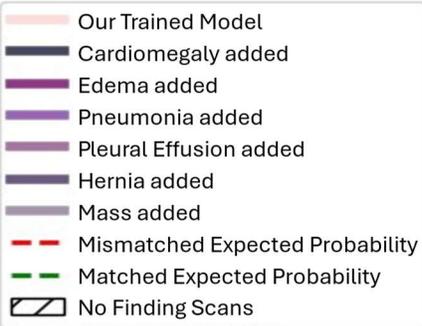

Supplemental Figure 6: Change in predicted probability of six conditions from baseline scans with no pathology to counterfactual scans with added pathology

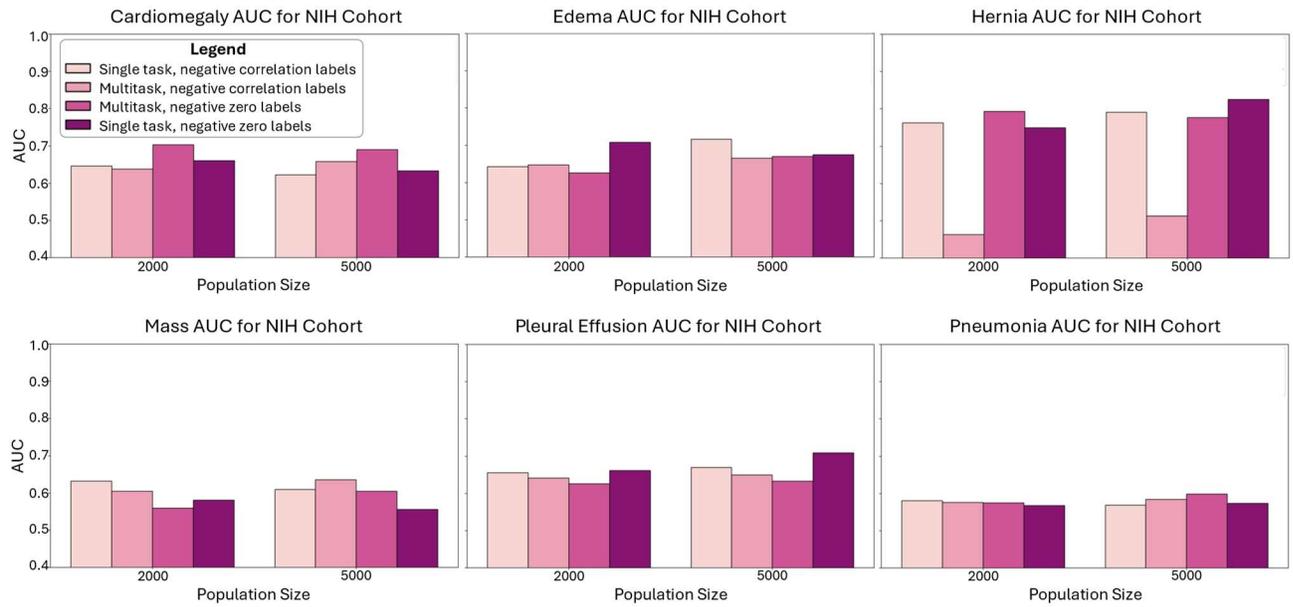

Supplemental Figure 7: small-scale diagnostic model AUC results to determine at scale diagnostic model training parameters